\documentclass[%
        final,
        notitlepage,
        narroweqnarray,
        inline,
        ]{ieee}
\usepackage{ieeefig}
\usepackage{amsfonts}
\usepackage{epsfig}
\usepackage{graphicx}
\usepackage{stfloats}

\begin{document}

\title[DPC for Fading Channels]{%
       Dirty Paper Coding for Fading Channels with
Partial Transmitter Side Information}

\author[C. S. Vaze and M. K. Varanasi]{%
   Chinmay S. Vaze and Mahesh K. Varanasi
    \thanks{%
This work was supported in part by NSF Grants CCF-0431170 and
CCF-0728955. The authors are with the Department of Electrical and
Computer Engineering, University of Colorado, Boulder, CO 80309-0425
USA (e-mail: {Chinmay.Vaze, varanasi}@colorado.edu, Ph:
001-303-492-7327).
      }
}

\maketitle

\begin{abstract}
The problem of Dirty Paper Coding (DPC) over the Fading Dirty Paper
Channel (FDPC) $Y=H(X+S)+Z$, a more general version of Costa's
channel, is studied for the case in which there is partial and
perfect knowledge of the fading process $H$ at the transmitter
(CSIT) and the receiver (CSIR), respectively. A key step in this
problem is to determine the optimal inflation factor (under Costa's
choice of auxiliary random variable) when there is only partial
CSIT. Towards this end, two iterative numerical algorithms are
proposed. Both of these algorithms are seen to yield a good choice
for the inflation factor. Finally, the high-SNR (signal-to-noise
ratio) behavior of the achievable rate over the FDPC is dealt with.
It is proved that FDPC (with $t$ transmit and $r$ receive antennas)
achieves the largest possible scaling factor of $\mathrm{min}(t,r)
\log \mathrm{SNR}$ even with no CSIT. Furthermore, in the high SNR
regime, the optimality of Costa's choice of auxiliary random
variable is established even when there is partial (or no) CSIT in
the special case of FDPC with $t \leq r$. Using the high-SNR
scaling-law result of the FDPC (mentioned before), it is shown that
a DPC-based multi-user transmission strategy, unlike other
beamforming-based multi-user strategies, can achieve a single-user
sum-rate scaling factor over the multiple-input multiple-output
Gaussian Broadcast Channel with partial (or no) CSIT.
\end{abstract}


\section{Introduction}
\PARstart In this paper, we study a more general version of Costa's
(original) Dirty Paper Coding (DPC) problem wherein present a fading
process, i.e., the problem of DPC over the channel of the form $Y =
H(X+S) +Z$, which we call the Fading Dirty Paper Channel (FDPC). We
study this problem for the case in which there is partial and
perfect knowledge of the fading process $H$ at the transmitter
(CSIT) and at the receiver (CSIR), respectively. Before continuing
with the problem at hand, we first explain the original DPC problem
studied by Costa.

Costa's work \cite{Costa} is based on the capacity formula of
Gelfand and Pinsker. They proved that the capacity of a discrete
memoryless channel $p(y |x ,s)$ with side information $S$ known
non-causally at the transmitter but not at the receiver is given by
\begin{equation}
C = \max_{p(u,x|s)} I(U;Y) - I(U;S), \label{GPformula}
\end{equation}
where $U$ is a finite-alphabet auxiliary random variable (RV)
\cite{G-P}. Costa used this formula for finding the capacity of the
channel $Y = X + S + Z$, where $X$ is the transmitted signal with
power constraint $E(X^2) \leq P$; interference $S$ is a zero-mean,
variance $Q$ Guassian RV ($S \sim \mathcal{N}(0,Q)$) and is assumed
to be known non-causally at the transmitter but not at the receiver;
and $Z \sim \mathcal{N}(0,Q)$ is the additive noise. With the
Gaussian input distribution and the choice of auxiliary RV as $U=X +
\alpha S$, where $X$ and $S$ are independent and the parameter
$\alpha$ is the inflation factor whose optimal value was found to be
$\frac{P}{P+N}$, Costa proved that the interference $S$ does not
result in any loss of capacity or $C = \frac{1}{2} \log (1+
\frac{P}{N})$. Costa named this technique of canceling the known
interference as Dirty Paper Coding (DPC).

The problem of current interest, i.e., the application of DPC to
FDPC with partial CSIT is of practical importance from the point of
view of studying the performance of DPC over the multiple-input
multiple-output (MIMO) Gaussian Broadcast Channel (BC) with partial
CSIT. The most challenging part of this problem is to find the
optimal inflation factor. Though this problem has been considered
before, the existing solutions are not satisfactory. In
\cite{Bennatan}, Bennatan et al. suggest a numerical approach which
involves an exhaustive search over a set which is arbitrarily
restricted to inflation factors that are optimal under perfect CSIT.
There is no reason for the inflation factor optimal under partial
CSIT to belong to this set. Moreover, such an exhaustive search can
be impractical or even impossible to implement. In \cite{Kotagiri},
Zhang et al. suggest the use of Costa's inflation factor (i.e.,
$\alpha = \frac{P}{P+N}$) over the SISO FDPC, as well. This choice
is clearly not optimal. Lastly the paper \cite{Piantanida} by
Piantanida et al. studies only a very specific setting of SISO FDPC
and therefore lacks generality. Thus the important problem of
determination of inflation factor still remains unresolved.

In this paper, we develop two algorithms for finding the inflation
factor under partial CSIT. These algorithms yield really good
results. Then, the paper deals with the high-SNR (signal-to-noise
ratio) analysis of the FDPC and some key results are proved on this
front.

\hspace{4 pt} \emph{\underline{Notation:}} An upper-case letter
(e.g., $X$) denotes a RV while the corresponding lower-case letter
(e.g., $x$) denotes its realization. $E_X(\cdot)$ denotes
expectation over RV $X$. For matrix $A$, $|A|$ denotes its
determinant while $\mathrm{tr}(A)$ is its trace, and its
complex-conjugate transpose is $A^*$. $I$ denotes the identity
matrix.

\section{Channel Model}   \label{ChannelModel}
A $t \times r$ FDPC is given by
\begin{equation}
Y = H ( X + S ) + Z.
\end{equation}
Here, the transmitted signal $X$ is a complex normal random vector
with mean $0$ and covariance matrix $\Sigma_X$ (i.e., $X \sim
\mathcal{C}\mathcal{N}(0, \Sigma_X)$) and has a power constraint of
$\mathrm{tr}(\Sigma_X) \leq P$; $S \sim \mathcal{C}\mathcal{N}(0,
\Sigma_S)$ is the interference known non-causally at the transmitter
but not at the receiver; $Z \sim \mathcal{C}\mathcal{N}(0,
\Sigma_Z)$ is the additive noise; and $X$, $S$, and $Z$ are assumed
to be independent. Channel fading matrix $H$ is assumed to be known
perfectly at the receiver whereas we let $\hat{H}$ to be the
transmitter's estimate of $H$. Further assume that $H$ is full rank
with probability $1$ and $|\Sigma_X|, |\Sigma_Z| > 0$. Let
$|\Sigma_S| = Q$, $|\Sigma_Z| = N$, and define signal-to-noise ratio
(SNR) to be $\frac{P}{N}$. We choose the auxiliary RV as $U = X +WS$
(Costa's choice extended to the MIMO FDPC), where the $t \times t$
matrix, $W$, is the inflation factor.

Using the capacity formula of \cite{Chiang} which is a
generalization of (\ref{GPformula}), we derive the achievable rate
over the FDPC with partial CSIT as
\begin{eqnarray}
\lefteqn{ R= E_{\hat{H}} \biggl( E_{H|\hat{H}} \log \hspace{1 pt} \{
\hspace{1 pt} |\Sigma_X| \hspace{1 pt} |H(\Sigma_X + \Sigma_S)
H^* + \Sigma_Z| \hspace{1 pt} \} - \biggr. }  \label{Rach} \\
&& \hspace{-6 pt} \left. \min_W E_{H|\hat{H}} \log \left| \hspace{-5
pt}
    \begin{array}{cc}
      \Sigma _X + W \Sigma _S W^* \hspace{-2pt}   & (\Sigma _X + W \Sigma _S) H^* \\
      H (\Sigma _X + \Sigma _S W^*) \hspace{-2pt} & H ( \Sigma _X + \Sigma _S ) H^* + \Sigma _Z \\
    \end{array}
 \hspace{-5 pt}  \right| \right).  \nonumber
\end{eqnarray}
Minimization over $W$ in the second term above is precisely the
problem of determination of inflation factor. We define the
no-interference upper-bound on the achievable rate as the rate
achievable over the FDPC in absence of interference (i.e., when
$Q=0$).

Under perfect CSIT, it is well established that the choice $U = X+ W
S$ is optimal and the optimal inflation factor is given by
$W_{opt}=\Sigma_X H^* (H \Sigma_X H^* + \Sigma_Z)^{-1} H$. However,
unlike the perfect-CSIT case, under partial CSIT, the choice $U = X
+ W S$ is not known to be optimal. Further, even if this choice of
auxiliary RV is assumed, the problem of finding a closed-form
solution for the optimal inflation factor appears intractable.

\section{Determination of Inflation Factor}
We first consider the case of SISO FDPC separately and then move on
to the general MIMO case.
\subsection{SISO FDPC $(t=r=1)$}
In the case of SISO FDPC (for which the inflation factor is scalar),
it is possible to generalize the known perfect-CSIT result to the
partial-CSIT case.
\newtheorem{proposition}{Proposition}
\begin{proposition} \label{PropW01}
For the SISO FDPC, the optimal inflation factor $W_{opt} \in [0,1]$,
irrespective of $H$ and $\hat{H}$.
\end{proposition}
\begin{proof}
In the SISO case, the problem reduces to
\[
\min_{W} E_{H|\hat{H}} \log \{|H|^2 P Q |1 - W | ^2 + |W| ^2 Q N + P
N\}.
\]
It can be seen that $W_{opt}$ must be real. Also, for any value of
$H=h$, function $f(W)= |h|^2 P Q |1 - W | ^2 + |W| ^2 Q N + P N$ is
quadratic in (real) $W$, and $W$ minimizing $f(W)$ lies in the
interval $[0,1]$ for any $h$.
\end{proof}
This proposition helps in numerical determination of $W_{opt}$. The
result of this proposition is quite surprising because such a result
can not be proved if the fading coefficients multiplying $X$ and $S$
are different.

\subsection{MIMO FDPC}
A result analogous to Proposition \ref{PropW01} can not be proved in
the MIMO case. Further, the minimization problem in (\ref{Rach}) is
a non-convex optimization problem. Also, the required conditional
expectations are analytically intractable for any general $H$ and
$\hat{H}$. This makes the problem difficult in the case of MIMO
FDPC. We now propose two suboptimal algorithms for finding the
inflation factor. The main advantage of our algorithms is that these
can be used over any general FDPC, irrespective of the distribution
of $H$ and the nature of partial CSIT $\hat{H}$.

\emph{Algorithm 1: } The key idea behind this algorithm is to
minimize the objective function stepwise, i.e., at each step, we
minimize over one row of $W$ while treating all other rows as
constants. It turns out that the objective function when regarded as
a function of only one row has a form that is amenable to analytical
closed-form minimization if we upper-bound it by moving the
expectation inside the logarithm. Thus, at every iteration,
minimization over each row of $W$ is carried out successively, and
these iterations are repeated until a good choice is found.

Let us now consider the minimization over the first row of $W$ while
treating all other rows of it as constants \footnote{Notation used
in Algorithm 1: For matrix $A$, $A_1$ denotes the first row of $A$,
$A^{(1)}$ denotes its first column, $A_{11}$ is its $(1,1)$ element,
and $A^{\bar{1}}$ is entire matrix $A$ except the first column of it
while $A_{\bar{1}}$ is entire matrix $A$ except the first row of it.
Thus, for example, $(W^*)^{\bar{1}}$ is the entire matrix $W^*$
except for first column of it; $(W^*)^{\bar{1}} = (W_{\bar{1}})^*$;
and $\Sigma_{X_1}^{\bar{1}}$ is part of first row of $\Sigma_X$
except for its $(1,1)$ element.}. Observe that only the first row
and the first column of the block-partitioned matrix in
(\ref{Rach}), which we call $M$, depend on $W_1$. Thus, we
repartition $M$, i.e., write $M = \left[
      \begin{array}{cc}
        a & B^* \\
        B & D\\
      \end{array}
    \right]$,
where $a= \Sigma_{X_{11}} + W_1 \Sigma_S (W_1)^*$ is $M_{11}$, $B^*
= [\Sigma_{X_1}^{\bar{1}} + W_1 \Sigma_S (W^*)^{\bar{1}}
\hspace{5pt} (\Sigma_{X_1} + W_1 \Sigma_S) H^*]$, and $D$ is a
$(t+r-1) \times (t+r-1)$ square matrix remained after excluding the
first row and the first column from $M$. Now, we have $|M|=|D|
|a-B^* D^{-1} B|$ \footnote{Matrix $D$ can be proved to be
invertible with probability $1$ for any choice of $W$.}. Since we
are minimizing only over $W_1$ while treating all other rows of $W$
as constants, $|D|$ will not affect the minimization. In order that
the required conditional expectations in the above minimization can
be computed (even) numerically, we upper-bound the objective
function using Jensen's Inequality. Thus we arrive at the following
optimization problem:
\begin{equation}
E_{\hat{H}} \log \min_{W_1} E_{H|\hat{H}} (a-B^* D^{-1} B).
\end{equation}
If $t=1$, we can directly obtain $W$ as given by equation
(\ref{W_t=1}) at the bottom of the next page, otherwise we proceed
as follows. We first evaluate the objective function above. Since
$B$ is in partitioned form, we partition $D^{-1}$ accordingly, i.e.,
let $D^{-1} = \left[
           \begin{array}{cc}
             F & G \\
             J & K \\
           \end{array}
         \right]$,
where $F$ is $(t-1) \times (t-1)$ matrix, $K$ is $r \times r$
matrix, and $G=J^*$. With this, we get equation (\ref{ab*d-1b}).
\begin{figure*}[!b]
\begin{picture}(10,1)
\put(10,10){\line(1,0){500}}
\end{picture}
\begin{eqnarray}
W = P E\big(H^* \{H(P+Q)H^*+\Sigma_Z\}^{-1} H \big) \{1-Q E \big(H^*
(H(P+Q)H^*+\Sigma_Z)^{-1} H \big) \} ^{-1}, \mbox{ where $E(\cdot)
\equiv E_{H|\hat{H}}(\cdot)$.} \label{W_t=1}
\end{eqnarray}
\begin{eqnarray}
\lefteqn{ a-B^* D^{-1} B = } \nonumber \\
&& {} \Sigma_{X_{11}} + W_1 \Sigma_S (W_1)^* - \big\{
(\Sigma_{X_1}^{\bar{1}} + W_1 \Sigma_S (W^*)^{\bar{1}} ) F
(\Sigma_{X_1}^{\bar{1}} + W_1 \Sigma_S (W^*)^{\bar{1}} )^* + (
\Sigma_{X_1} + W_1 \Sigma_S) H^* J (\Sigma_{X_1}^{\bar{1}} + W_1 \Sigma_S (W^*)^{\bar{1}} )^* \big. +  \nonumber \\
&& {} \big. (\Sigma_{X_1}^{\bar{1}} + W_1 \Sigma_S (W^*)^{\bar{1}} )
G H (\Sigma_{X_1} + W_1 \Sigma_S)^* + (\Sigma_{X_1} + W_1 \Sigma_S)
H^* K H (\Sigma_{X_1} + W_1 \Sigma_S)^* \big\}. \label{ab*d-1b}
\end{eqnarray}
\begin{eqnarray}
\lefteqn{ \hspace{-14cm} W_1 = \left( \Sigma_{X_1}^{\bar{1}} E(F)
W_{\bar{1}} \Sigma_S + \Sigma_{X_1} E(H^*J) W_{\bar{1}} \Sigma_S +
\Sigma_{X_1}^{\bar{1}} E(GH) \Sigma_S + \Sigma_{X_1} E(H^*K H)
\Sigma_S\right) \times \left( \Sigma_S -\Sigma_S (W_{\bar{1}})^*
E(F) W_{\bar{1}} \Sigma_S - \right. }\nonumber
\\ \hspace{-3cm} \hspace{4pt} \left. \Sigma_S E(H^*J) W_{\bar{1}} \Sigma_S - \Sigma_S
(W_{\bar{1}})^* E(GH) \Sigma_S - \Sigma_S E(H^*KH) \Sigma_S \right)
^{-1}, \mbox{ where $E(\cdot) \equiv E_{H|\hat{H}}(\cdot)$.}
\label{W_1opt}
\end{eqnarray}
\end{figure*}
It can be observed that the expression $a-B^* D^{-1} B$ is quadratic
in $W_1$. Thus, using technique of `completing the square,' one can
evaluate the optimal $W_1$ and is given by equation (\ref{W_1opt})
\footnote{It turns out that the matrix inverted in (\ref{W_1opt}) is
invertible if and only if $\Sigma_S$ is invertible. This can be
easily fixed via spectral decomposition of $\Sigma_S$. Due to lack
of space, we omit details of it here.}. Here, we need conditional
expectations of $F$, $GH$ (or $H^*J$), and $H^* K H$, which can be
evaluated numerically through Monte-Carlo Simulations.

Now, consider minimization over $W_k$, treating all other rows of
$W$ as constants. Similar to the case of $W_1$, observe that only
the $k^{th}$ row and the $k^{th}$ column of $M$ depend on $W_k$.
Thus, by interchanging the $k^{th}$ row of $M$ with its first row
and the $k^{th}$ column with the first column, we can obtain a
matrix $M'$ whose only first row and the first column depend on
$W_k$. Hence, we can use the procedure described before to minimize
$|M'| = |M|$ over $W_k$.

An iterative algorithm can now be set up as follows:
\begin{enumerate}
  \item Start with some initial choice for $W$. We typically use $W=I$
  for this purpose.
  \item Loop: $k=1$ to t
  \begin{itemize}
    \item Minimize over $W_k$ using the procedure described before.
    \item Update $W$ according to $W_k$ found before and so also
    matrix $M$.
  \end{itemize}
  \item Repeat Step 2 until the increase in the achievable rate is
  negligible.
\end{enumerate}
The algorithm produces a bounded (from below), monotonically
decreasing sequence of upper-bounds on the objective function with
iteration steps and hence, it converges.

\newtheorem{simulation}{Simulation}
\begin{simulation}
For simplicity we take the elements of $H$ to be independent and
identically distributed as $\mathcal{N}(0,1)$. We quantize each
element separately using a simple equal spacing level quantizer
(quantization bins are of equal length except for the first and last
bins that extend to infinity) and the spacing is determined using
data from \cite{Max}. In Fig. 1(a), $B$ denotes the number of
feedback bits per element of matrix $H$.

Fig. 1(a) plots the achievable rate for the $3 \times 2$ and $3
\times 3$ MIMO FDPCs as a function of $P$. Comparing $R$ with the
no-interference upper-bound, it can be said that our algorithm finds
a good choice for the inflation factor. Also, some important
observations can be made regarding the high-SNR behavior of the
achievable rate. These observations have been formally stated and
proved as Theorems 1 and 2 later in this paper.
\end{simulation}

\begin{figure*}[!b]
\centering \hspace{-30pt}
\includegraphics[height=2.8in,width=7in]{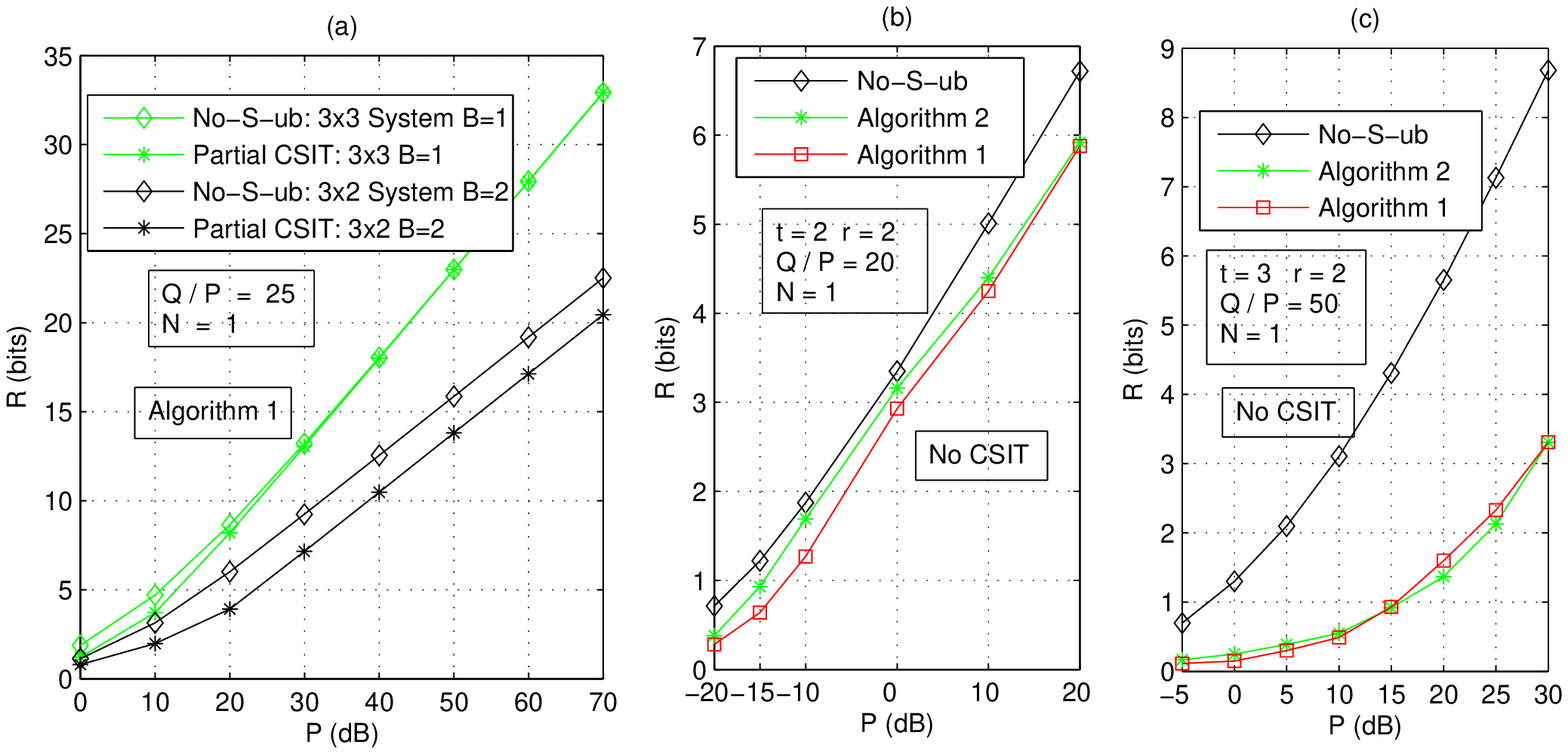}
\caption{Numerical Results Using Two Algorithms. (a) Achievable
Rates vs. P: Algorithm 1. (b) Comparison of Two Algorithms: $2
\times 2$ FDPC. (c) Comparison of Two Algorithms: $3 \times 2$ FDPC.
(No-S-ub denotes no-interference upper-bound.) }
\end{figure*}

\emph{Algorithm 2: } In the previous algorithm, we minimize the
upper-bound (obtained via Jensen's Inequality) on the objective
function. The key idea of this second algorithm is to solve the
Karush-Kuhn-Tucker conditions for the optimization problem or in
other words, to find a stationary point of the objective function.

Thus, we need to solve an equation $\frac{d}{dW} E_H \log |M| =0$,
where $M$ is the block-partitioned matrix in (\ref{Rach}) as
referred in the derivation of Algorithm 1. For simplicity, we
consider in this derivation only the case of no CSIT since the
results can be easily extended to the partial-CSIT case. We start
with finding differential \footnote{Since the matrices involved in
the optimization are all complex-valued, we basically need to
convert the problem to an equivalent optimization problem involving
only real-valued variables using the technique of \cite{Telatar}.
However, due to lack of space, we consider in this derivation only
the case of real-valued variables. It can be proved that even in the
case of complex-valued variables the equation (\ref{Algo2finaleqn})
still holds.} as follows:
\begin{eqnarray*}
\lefteqn{ d E_H \log |M|  =  E_H \mathrm{tr} \{M^{-1} dM\} \hspace{5pt} \cdots \mbox{ see \cite{Magnus}. }}\\
&& {} =  E_H \mathrm{tr} \left\{\hspace{-1pt} M^{-1} \hspace{-2pt}
\left[ \hspace{-3pt}
\begin{array}{cc}
dW \Sigma_S W^* + W \Sigma_S (dW)^* \hspace{-3pt} & (dW) \Sigma_S H^*\\
H \Sigma_S (dW)^* \hspace{-3pt} & 0 \\
\end{array}
\hspace{-3pt} \right] \hspace{-1pt} \right\} \\
&& {} =  2 \hspace{1pt} E_H \mathrm{tr} \left\{ \Sigma_S \left[
\begin{array}{cc}
W^* & H^*
\end{array} \right] M^{-1}
\left[
\begin{array}{c}
dW \\
0 \\
\end{array}
\right] \right\}.
\end{eqnarray*}
Thus we arrive at an equation $\frac{d}{dW} E_H \log |M| =0
\Rightarrow$
\begin{equation}
E_H (A_1W+A_2^*H)\Sigma_S = 0, \mbox{ where} \left[ \hspace{-2pt}
\begin{array}{c}
A_1 \\
A_2 \\
\end{array}\hspace{-2pt}
\right] \hspace{-2pt} = M^{-1} \left[ \hspace{-2pt}
\begin{array}{c}
    I \\
    0 \\
\end{array} \hspace{-2pt}
\right]. \label{Algo2finaleqn}
\end{equation}
It can be proved that without loss of generality we can consider a
solution of the form $W=-(E_H A_1)^{-1} E_H(A_2^*H) = -g(W)$ for the
above equation\footnote{Matrix $A_1$ is invertible with probability
$1$.} even when $|\Sigma_S|=0$. This fixed-point equation for $W$
allows us to set up the following iterative algorithm.
\begin{enumerate}
  \item Start with some initial choice for $W$, i.e., $W^{(0)}$.
  \item At the $n^{th}$ iteration, set $W^{(n)}= -g(W^{(n-1)})$ with
  required expectations computed numerically. \label{Step2algo2}
  \item Repeat Step \ref{Step2algo2} until the improvement in the
  achievable rate is negligible.
\end{enumerate}
Since our optimization problem is non-convex, the algorithm does not
necessarily yield the optimum solution.

\begin{simulation}
It is observed, rather surprisingly, that in most of the cases, both
the algorithms yield $W$s that achieve almost equal rate over the
FDPC. Thus, rather than considering such cases, we include here some
of the plots in which one algorithm does better than the other in
terms of the achievable rate. Figs. 1(b) and 1(c) show some of these
cases. However, it has not been possible to precisely characterize
such cases. The main advantage of Algorithm 1 over the second is
that Algorithm 1 is usually faster in terms of number of iterations
required.
\end{simulation}

\begin{figure}
\hspace{-13pt}
\includegraphics[height=1.15in,width=3.7in]{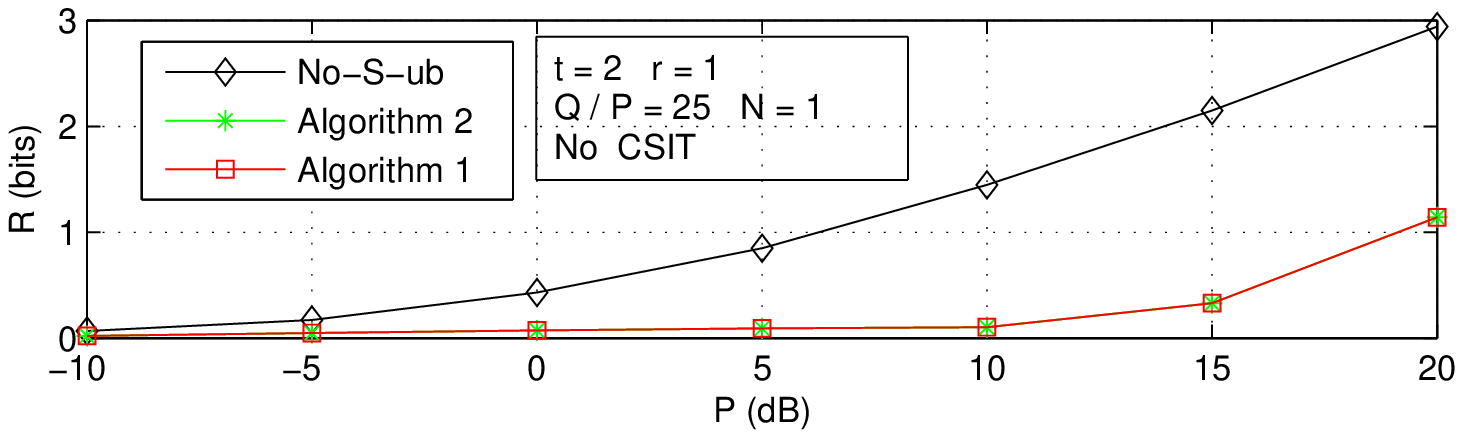}
\caption{A Case in which Algorithms Do Not Work.} \label{problemfig}
\end{figure}

It would be in order here to acknowledge a limitation of these
algorithms. It is observed sometimes that for systems with $t
>r$, these algorithms do not yield a good solution.
One such example is seen in Fig. \ref{problemfig} in which it seems
likely to us that some improvement in R is possible in a range of
$P$ = $0$ to $10$ dB. It has been puzzling that even though these
algorithms solve for two different things (one minimizes an
upper-bound on the objective function while other solves for the
stationary point of it), both of them fail at the same time and find
nearly same solutions (in terms of R).

\emph{Application to the lattice-based transmission strategies:} A
lattice-based transmission scheme for the MIMO FDPC with no CSIT has
been proposed in \cite{SCLin}. The problem of determination of
inflation factor also appears in the design of such a scheme, for
which the paper \cite{SCLin} does not provide any satisfactory
solution. Thus, our algorithms can be used for the design of
lattice-based schemes as well.

\section{High SNR Analysis}  \label{HighSNR}
In this section, we deal with the high-SNR behavior of the
achievable rate and prove some important results.

\subsection{FDPC}
We begin with the high-SNR scaling law of the FDPC.
\newtheorem{theorem}{Theorem}
\begin{theorem} \label{ScalingLawFDPC}
The rate achievable over the no (and hence partial) CSIT FDPC using
DPC scales optimally in the high-SNR regime as $\mathrm{min}(r,t)
\log \mathrm{SNR}$ if the ratio $\frac{Q}{P}$ is held constant
\footnote{It is relevant from the point of view of MIMO Gaussian BC
to consider $\frac{Q}{P}$ remaining constant as $P \to \infty$ and
hence the same assumption is made in earlier simulations as well.}
as $P \to \infty$.
\end{theorem}
\begin{proof}
It is clear that the scaling factor of FDPC with partial or no CSIT
is upper-bounded by $\mathrm{min}(r,t) \log \mathrm{SNR}$. We, in
fact, prove that it is equal to $\mathrm{min}(r,t) \log
\mathrm{SNR}$ by proving that a lower-bound on the achievable rate
can achieve the before-mentioned scaling factor.

Let us consider a lower-bound obtained by using a particular choice
of $W=I$.
\begin{eqnarray}
R & \geq & E_H \log \frac{|\Sigma_X|
|\Sigma_Z+H(\Sigma_X+\Sigma_S)H^*|} { \left|
                    \begin{array}{cc}
                        \Sigma_X + \Sigma_S & (\Sigma_X + \Sigma_S)H^* \\
                        H(\Sigma_X+\Sigma_S) & H(\Sigma_X + \Sigma_S)H^*+\Sigma_Z \\
                    \end{array}
                    \right| } \nonumber \\
& = & E_H \log \frac{|\Sigma_X| |\Sigma_Z+H(\Sigma_X+\Sigma_S)H^*|}
{ \left|
\begin{array}{cc}
    \Sigma_X + \Sigma_S & 0 \\
    0 & \Sigma_Z \\
\end{array}
\right| } \label{equality1}\\
& \geq & E_H \log \left\{ \frac{|\Sigma_X|}{|\Sigma_X+\Sigma_S|}
\frac{|\Sigma_Z + H \Sigma_X H^*|}{|\Sigma_Z|} \right\}
\label{inequality2},
\end{eqnarray}
where the equality in (\ref{equality1}) follows due to row and
column operations, whereas the inequality in (\ref{inequality2})
follows because $\Sigma_X+\Sigma_S \succeq \Sigma_X$ within partial
order.

Since $\frac{Q}{P}$ is assumed to remain constant as $P \to \infty$,
the first factor $\frac{|\Sigma_X|}{|\Sigma_X+\Sigma_S|}$ remains
bounded as $P \to \infty$. The second factor $E_H \log
\frac{|\Sigma_Z + H \Sigma_X H^*|}{|\Sigma_Z|}$ can be made to scale
in the high-SNR regime as $\mathrm{min}(r,t) \log \mathrm{SNR}$,
even without any CSIT, by choosing an appropriate $\Sigma_X$.
\end{proof}
The theorem above proves the achievability of the largest possible
scaling factor over the partial or no CSIT MIMO FDPC and in this
sense, the statement of the theorem is the strongest.

The above theorem can be further strengthened in the case of FDPC
with $t \leq r$.
\begin{theorem}
For the FDPC with $t \leq r$, if the ratio $\frac{Q}{P}$ is held
constant, the difference $\triangle R$ between the no-interference
upper-bound and the rate achievable under no (and hence partial)
CSIT tends to zero as $P \to \infty$.
\end{theorem}
\begin{proof}
Under the specific choice of $W=I$, we get
\begin{eqnarray}
\lefteqn{ \triangle R = E_H \log \frac{|\Sigma_Z + H \Sigma_X
H^*|}{|\Sigma_Z + H(\Sigma_X + \Sigma_S)H^*| } - \log
\frac{|\Sigma_X|}{|\Sigma_X + \Sigma_S|} }\nonumber \\
&& {} = E_H \log \frac{|I + T H \Sigma_X H^* T^*|} { |I + T H
(\Sigma_X + \Sigma_S) H^* T^*|} - \log
\frac{|\Sigma_X|}{|\Sigma_X + \Sigma_S|} \nonumber \\
&& {} = E_H \log \frac{|P^{-1} I + \Sigma_X' H^* \Sigma_Z^{-1} H|}{
|P^{-1} I + (\Sigma_X' +\Sigma_S') H^* \Sigma_Z^{-1} H|} - \log
\frac{|\Sigma_X'|}{|\Sigma_X' + \Sigma_S'|}, \nonumber
\end{eqnarray}
where $\Sigma_Z = T^{-1}(T^*)^{-1}$, $\Sigma_X= P \Sigma_X'$,
$\Sigma_S = P \Sigma_S'$, and the third equality follows from the
fact that $|I+AB| = |I+BA|$ whenever products $AB$ and $BA$ are
defined.

For the case of $t \leq r$, matrix $H^* \Sigma_Z^{-1} H$ is
invertible. Thus, we obtain $\lim_{P \to \infty} \triangle R = 0$.
Therefore, the choice $W=I$ is optimal at high SNR for FDPCs with $t
\leq r$.
\end{proof}
For $t > r$ case, $\triangle R$ does not indeed to go to zero,
except in some special cases.

There is a nice intuitive explanation of why $\triangle R$ tends to
zero when $t \leq r$. Consider the perfect-CSIT-optimal $W$, i.e.,
$W= \Sigma_X H^* (H \Sigma_X H^* +\Sigma_Z)^{-1} H$. For $t \leq r$,
it can be proved that $W \to I$ as $P \to \infty$ for any value of
$H$. Hence, when $t \leq r$, in the limit of high SNR, knowledge of
$H$ is not required as far as determination of $W$ is concerned, and
therefore, we get $\triangle R = 0$ in limit. However, for the case
of $t>r$, we get $W \to \Sigma_X' H^*(H \Sigma_X' H^*)^{-1} H$ (as
$P \to \infty$) which depends on the value of $H$. Hence, for FDPCs
with $t> r$, even in the limit, CSIT is required for determination
of $W$, and thus $\triangle R$ does not to go to zero in general.

An important implication of the above theorem is the following
proposition.
\begin{proposition}
For FDPCs with $t \leq r$, Costa's choice of auxiliary RV, namely,
$U=X+WS$ is optimal at high SNR even under partial or no CSIT.
\end{proposition}
Thus, for the first time the optimality of Costa's choice of
auxiliary RV is shown under partial CSIT (even though it is only in
a special case).

\subsection{Gaussian MIMO BC}
Theorem \ref{ScalingLawFDPC} proved in the previous subsection has
an important consequence on the achievable sum-rate scaling factor
of the Gaussian MIMO BC with partial or no CSIT. It is now
well-established that given a fixed level of partial CSIT
beamforming-based multi-user transmission strategies can not achieve
sum-rate scaling over the MIMO BC \cite{Jindal}. And thus, till now,
only the single-user strategy of time-division multiple access
(TDMA) was known to achieve to the sum-rate scaling. However, as the
following proposition suggests, a DPC-based multi-user transmission
strategy can also achieve the same.
\begin{proposition}   \label{PropBC}
For the Gaussian MIMO BC with $t$ transmit antennas and users with
$r$ receive antenna each, a high-SNR sum-rate scaling factor of
$\mathrm{min}(t,r) \log \mathrm{SNR}$ can be achieved even without
any CSIT if DPC is used.
\end{proposition}
\begin{proof}
If DPC is used at the transmitter, for the user encoded last, unlike
other users, entire interference can be (potentially) canceled by
DPC. Hence, as per Theorem \ref{ScalingLawFDPC}, the achievable rate
for the last user can be made to scale in the high-SNR regime as
$\mathrm{min}(t,r) \log \mathrm{SNR}$.
\end{proof}
Though TDMA can achieve the same scaling factor, this result is
interesting because it is for the first time that a multi-user
strategy is shown to achieve a non-zero high-SNR sum-rate scaling
factor over the Gaussian MIMO BC with partial or no CSIT. Also, this
proposition is in accordance with the main result and the conjecture
of \cite{Lapidoth}.

\section{Conclusion and Future Scope}
The paper proposes two good algorithmic solutions for the problem of
determination of inflation factor. Apart from this, some important
results are proved analytically in the high SNR regime. Our
algorithmic solutions are found to work well, except in some cases
as mentioned before. More efforts are required to better the
performance in these cases.

\bibliographystyle{IEEEbib}
\bibliography{final}

\begin{thebibliography}{10}

\bibitem{Costa}
M.Costa,
\newblock ``Writing on dirty paper,''
\newblock {\em IEEE Trans. Inform. Theory}, vol. 29, no. 3, pp. 439--441, May
  1983.

\bibitem{G-P}
S.Gelfand and M.Pinsker,
\newblock ``Coding for channel with random parameters,''
\newblock {\em Problems of control and information theory}, vol. 9, no. 1, pp.
  19--31, 1980.

\bibitem{Bennatan}
A.~Bennatan and D.~Burshtein,
\newblock ``On the fading paper achievable region of the fading {M}{I}{M}{O}
  broadcast channel,''
\newblock in {\em 44th annual Allerton Conference on Communication, Control and
  Computing}, Univ. of Illinois, Sep. 2006.

\bibitem{Kotagiri}
W.Zhang, S.~Kotagiri, and J.~N. Laneman,
\newblock ``Writing on dirty paper with resizing and its application to
  quasi-static broadcast channel,''
\newblock in {\em IEEE Int. Symp. Information Theory}, Nice, France, Jun. 2007,
  pp. 381--385.

\bibitem{Piantanida}
P.~Piantanida and P.~Duhamel,
\newblock ``Dirty paper coding without channel information at the transmitter
  and imperfect estimation at the receiver,''
\newblock in {\em IEEE Int. Conf. Communications}, Glasgow, Scotland, Jun.
  2007, pp. 5406--5411.

\bibitem{Chiang}
T.~Cover and M.~Chiang,
\newblock ``Duality between channel capacity and rate distortion with two-sided
  state information,''
\newblock {\em IEEE Trans. Inform. Theory}, vol. 48, no. 6, pp. 1629--1638,
  Jun. 2002.

\bibitem{Max}
J.~Max,
\newblock ``Quantizing for minimum distortion,''
\newblock {\em IEEE Trans. Inform. Theory}, vol. 6, no. 1, pp. 7--12, Mar.
  1960.

\bibitem{Telatar}
E.~Telatar,
\newblock ``Capacity of multi-antenna gaussian channels,''
\newblock {\em European Trans. on Telecomm.}, vol. 10, pp. 585--595, 1999.

\bibitem{Magnus}
J.~Magnus and H.~Neudecker,
\newblock {\em Matrix Differential Calculus with Applications in Statistics and
  Econometrics},
\newblock John Wiley and Sons, Inc., 1988.

\bibitem{SCLin}
S.~C. Lin, P.~H. Lin, and H.~Su,
\newblock ``Lattice coding for the vector fading paper problem,''
\newblock in {\em IEEE Inform. Theory Workshop}, Tahoe City, CA, USA, Sep.
  2007, pp. 78--83.

\bibitem{Jindal}
N.~Jindal,
\newblock ``M{I}{M}{O} broadcast channels with finite rate feedback,''
\newblock {\em IEEE Trans. Inform. Theory}, vol. 52, no. 11, pp. 5045--5060,
  Nov. 2006.

\bibitem{Lapidoth}
A.~Lapidoth, S.~Shamai (Shitz), and M.~Wiger,
\newblock ``On the capacity of a {M}{I}{M}{O} fading broadcast channel with
  imperfect transmitter side-information,''
\newblock in {\em 43rd annual Allerton Conference on Communication, Control and
  Computing}, Monticello, IL, Sep. 2005.

\end{thebibliography}

\end{document}